\documentclass[conference]{IEEEtran}
\IEEEoverridecommandlockouts
\usepackage{amsmath,amssymb,amsfonts}
\usepackage{algorithmic}
\usepackage{graphicx}
\usepackage{textcomp}
\usepackage{xcolor}
\usepackage{float} 

\begin{document}

\title{Real-time CARFAC Cochlea Model Acceleration on FPGA for Underwater Acoustic Sensing Systems
}

\author{
\IEEEauthorblockN{Bram Bremer}
\IEEEauthorblockA{\textit{International Centre for Neuromorphic Systems} \\
\textit{Western Sydney University, Sydney, Australia}\\
a.bremer@westernsydney.edu.au}
\and
\IEEEauthorblockN{Matthew Bigelow}
\IEEEauthorblockA{\textit{Department of Defence} \\
\textit{DST Group, Australia} \\
matthew.bigelow@defence.gov.au}
\and
\IEEEauthorblockN{Stuart Anstee}
\IEEEauthorblockA{\textit{Department of Defence} \\
\textit{DST Group, Australia} \\
stuart.anstee@defence.gov.au}
\and
\IEEEauthorblockN{Gregory Cohen}
\IEEEauthorblockA{\textit{International Centre for Neuromorphic Systems} \\
\textit{Western Sydney University, Sydney, Australia }\\
G.cohen@westernsydney.edu.au}
\and
\IEEEauthorblockN{André van Schaik}
\IEEEauthorblockA{\textit{International Centre for Neuromorphic Systems} \\
\textit{Western Sydney University, Sydney, Australia }\\
A.VanSchaik@westernsydney.edu.au} 

\and
\IEEEauthorblockN{Ying Xu}
\IEEEauthorblockA{
\makebox[\textwidth]{\textit{International Centre for Neuromorphic Systems}}\\
\makebox[\textwidth]{\textit{Western Sydney University, Sydney, Australia}}\\
\makebox[\textwidth]{ying.xu@westernsydney.edu.au}
}
}
\maketitle
\begin{abstract}
This paper presents a real-time, energy-efficient embedded system implementing an array of Cascade of Asymmetric Resonators with Fast-Acting Compression (CARFAC) cochlea models for underwater sound analysis. Built on the AMD Kria KV260 System-on-Module (SoM), the system integrates a Rust-based software framework on the processor for real-time interfacing and synchronization with multiple hydrophone inputs, and a hardware-accelerated implementation of the CARFAC models on a Field-Programmable Gate Array (FPGA) for real-time sound pre-processing. Compared to prior work, the CARFAC accelerator achieves improved scalability and processing speed while reducing resource usage through optimized time-multiplexing, pipelined design, and elimination of costly division circuits. Experimental results demonstrate 13.5\% hardware utilization for a single 64-channel CARFAC instance and a whole board power consumption of 3.11 W when processing a 256 kHz input signal in real time.
\end{abstract}

\begin{IEEEkeywords}
CARFAC, Cochlea Model, Neuromorphic Engineering, FPGA, Accelerator, Embedded, Rust, Underwater Acoustic Sensing
\end{IEEEkeywords}

\section{Introduction}
Sound analysis in real-world environments presents significant challenges. In particular,  underwater sound source detection, classification, and localization are often hindered by factors such as transmission loss, multipath propagation, reverberation, and ambient noise \cite{Urick1983}. 
To improve the robustness of underwater acoustic signal analysis, multi-sensor systems--such as hydrophone arrays--are often used together with beamforming techniques to enable efficient spatial sampling of the acoustic environment. For example, the delay-and-sum beamformer combines sensor signals with relative time delays or phase shifts to reinforce sound energy from a particular direction. The Minimum Variance Distortionless Response (MVDR) beamformer, also known as Capon's method \cite{Capon1969}, and the Multiple Signal Classification (MUSIC)  \cite{LiDeepLearning}  exploit statistical properties of array data for high-resolution direction-of-arrival (DOA) estimation. These methods form the basis of many underwater sensing applications, including localization, navigation, and surveillance. 

\begin{figure*}[t]
    \centering
    \includegraphics[width=0.9\linewidth]{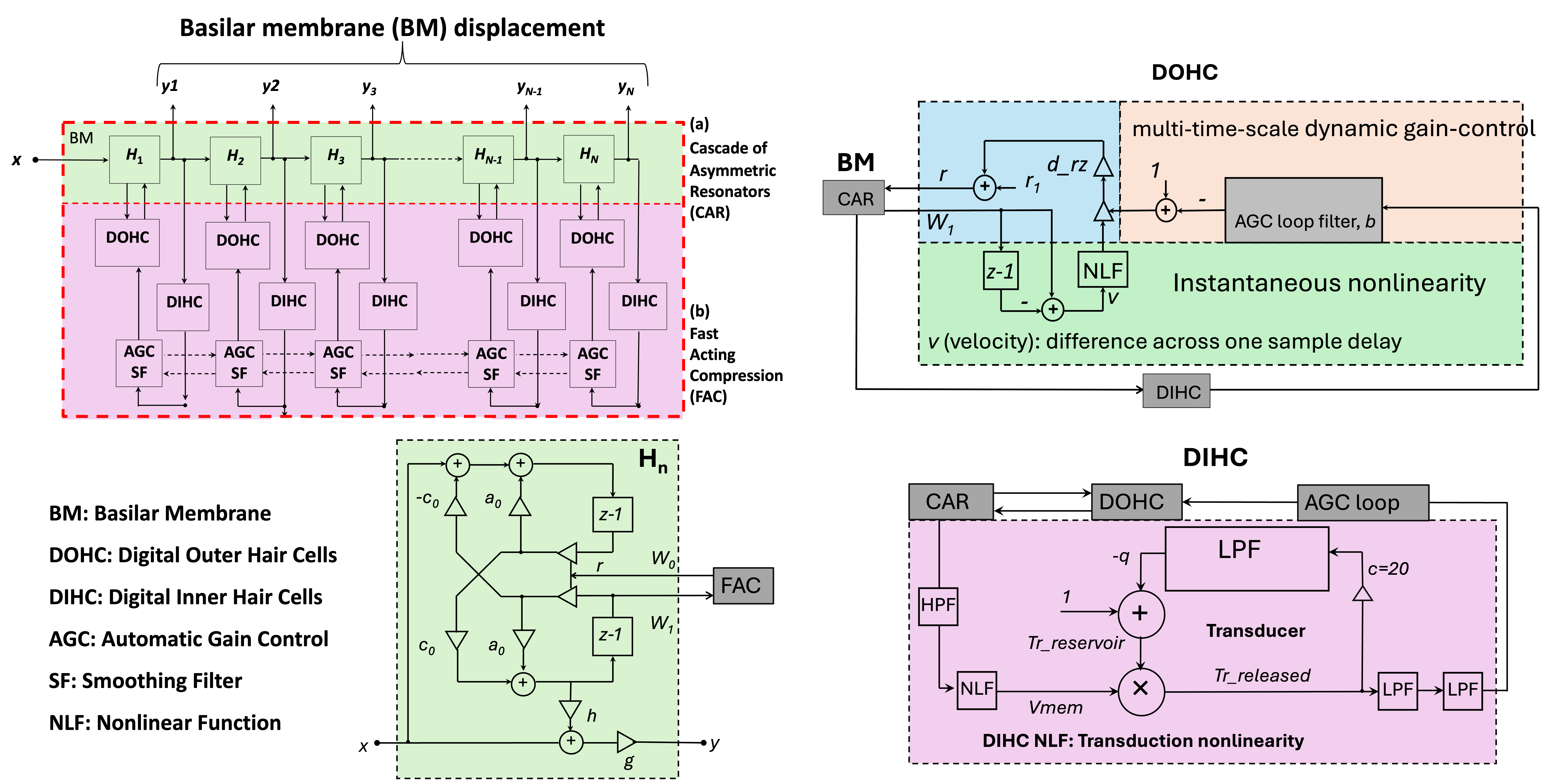}
    \caption{(Left top) Structure of the CAR-FAC model. \textit{x} is the input sound, $H_1$ to $H_N$ are the transfer functions of the CAR part, and $y_1$ to $y_N$ represent the CAR-FAC output. The best frequencies of the CAR resonators decrease from left to right. The DOHC, the DIHC, and the AGC loop comprise the FAC part. (Left bottom) Structure of one section of CAR, the two-pole-two-zero resonator, $H_n$. $a_0$, $c_0$, and $h$ are the resonator coefficients, $r$ is the pole/zero radius in the $z$ plane, $g$ is the DC gain factor, $W_0$ and $W_1$ are the intermediate variables, $x$ is the input, and $y$ is the output. (Right top) Structure of the DOHC. The instantaneous nonlinearity performs a nonlinear gain control on the CAR velocity, $v$ which is calculated from the BM coefficient $W_1$. The multi-time-multi-scale dynamic gain-control factor, $b$, is obtained from the AGC loop. Both gain control factors are combined to change $r$. (Right bottom) Structure of the DIHC \cite{Xu2018AModel}. The details of the CARFAC model can be found in  \cite{Lyon2017HumanHearing}.}    
    \label{fig:SimplCARFAC}
\end{figure*}

Beamforming approaches conventionally use short-time Fourier analysis directly as the front end, whereas an emerging alternative is the use of bio-inspired hearing techniques that mimic the efficiency of humans and animals in perceiving sound in complex acoustic environments. For example, Ni et al. \cite{rs16163074} and Guo et al. \cite{9849785} investigated gammatone filters and gammatone-based features combined with convolutional neural networks for underwater sound recognition; Maymon et al. \cite{AuditoryFilter} investigated the use of gammatone filters together with the MUSIC beamformer for sound localization in a reverberant environment, where the head-related transfer function (HRTF) was used to construct a steering vector.
Such filters provide bio-realistic spectral decomposition, but they are limited in dynamic acoustic environments because they consist of a linear, time-invariant filter bank, which lacks the adaptive gain control that the real cochlea exhibits. 
An alternative is to use cochlea models to perform time-frequency analysis more efficiently and adaptively, such as the CARFAC cochlea model.

The CARFAC simulates cochlear processing through a cascade of asymmetric resonators coupled with a fast automatic gain control (AGC) mechanism. 
It exhibits level-dependent frequency tuning, where the effective bandwidth and gain adapt to input levels. This level-dependent tuning mechanism has the potential to enhance signal representation under masking conditions.
The fast dynamic range compression in CARFAC preserves the sensors’ dynamic range when exposed to underwater explosive sound or ship noise. 
The time-domain expression of the CARFAC model preserves the fine temporal structure with instantaneous phase information, which is often lost in frame-based Fourier analysis. With realistic group delays per frequency channel, CARFAC outputs resemble simulated auditory nerve firing patterns, capturing both envelope and fine timing cues essential for localization and identification. 
These characteristics contribute to robustness against noise, nonstationarity, and multipath effects, making CARFAC a good candidate for underwater signal preprocessing. CARFAC has the potential to be integrated with frequency- or time-domain beamforming for power-efficient underwater sensing.

Autonomous underwater systems are often constrained by limited computational resources, making energy-efficient, real-time processing a key requirement. However, real-time sound localization and tracking impose significant computational demands. 
To address this challenge and investigate the potential of CARFAC in underwater acoustic sensing, we developed a real-time CARFAC-based preprocessing system for a multi-hydrophone array, implemented on the AMD Kria KV260 SoM. 
In our previous work in \cite{Xu2016ElectronicFPGA, Xu2018AModel}, we implemented the CARFAC model on FPGA platforms, and in \cite{XuLocalisation}, we applied the system to sound localization and demonstrated improved bearing accuracy. In this work, we improve the model for hardware implementation and optimize time-multiplexing and pipelining design to make it suitable for processing high-sampling-rate underwater signals. The system is designed for deployment on embedded platforms such as autonomous underwater vehicles (AUVs).
\\
\section{Method}
The CARFAC cochlea model was developed by Lyon in \cite{Lyon2017HumanHearing, Lyon2024TheJAX}. It approximates the physiological components of the human cochlea and mimics its qualitative behavior. It can be divided into four distinct sections based on their biological origin: the basilar membrane (represented as $H$), the inner and outer hair cells (IHC and OHC, respectively), and the medial olivocochlear efferent system, modeled by the AGC, as shown in Fig \ref{fig:SimplCARFAC}. In this work, we optimize the computational CARFAC model for efficient hardware implementation:

\subsection{Division Replacement}
The CARFAC model has three division operations. Efficient hardware implementation of division is challenging, often requiring iterative algorithms like Newton-Raphson to approximate the result. To accelerate the hardware design and reduce resource utilization, we approximate these division operations numerically using alternative expressions.
The first division operation is the non-linear function (NLF) component of the DIHC, $v_{\text{mem}}$, as shown in Fig. \ref{fig:SimplCARFAC} :

\begin{equation}
v_{\text{mem}} = \frac{p^3}{p^3 + p^2 + 0.1}
\label{eq:nlf_ihc}
\end{equation}

\begin{equation}
p = \max\left(0, \mathrm{BM\_hpf}[:, t] + 0.175\right)
\label{eq:u}
\end{equation}
where $v_\text{mem}$ is the DIHC membrane conductance as a function of $p$. $p$ is the half-wave rectifier with an offset, 0.175, as a function of the high pass filter out of all the BM channels, \text{BM\_hpf}[:,t] at time $t$, and as shown in Fig. \ref{fig:SimplCARFAC} (left bottom), the transfer function of each BM channel in the $z$ domain is:
\begin{equation}
\frac{Y}{X} = g\left[1 + h \cdot \frac{z c_0 r}{z^2 - 2 a_0 r z + \left(\left(a_0 r\right)^2 + \left(c_0 r\right)^2\right)}\right]
\end{equation}
where $X$ and $Y$ are the input and output of each BM channel. $a_0$, $c_0$, and $h$ are the resonator coefficients, $r$ is the pole/zero
radius in the $z$ plane, $g$ is the DC gain factor,
 In our implementation, $v_\text{mem}$ is modeled by: 
\begin{equation}
v_{\text{mem}} = 0.75 \cdot (1 - p)^2
\label{eq:vz1}
\end{equation}
\begin{equation}
p = \min\left(1,\ p_{\text{int}}^8\right)
\label{eq:z1}
\end{equation}
\begin{equation}
p_{\text{int}} = \max\left(0,\ 1 - \frac{\mathrm{BM\_hpf}[:,t] + 0.13}{4}\right)
\label{eq:z}
\end{equation}
\begin{figure}
    \centering
    \centerline{\includegraphics[height=4.6cm, keepaspectratio]{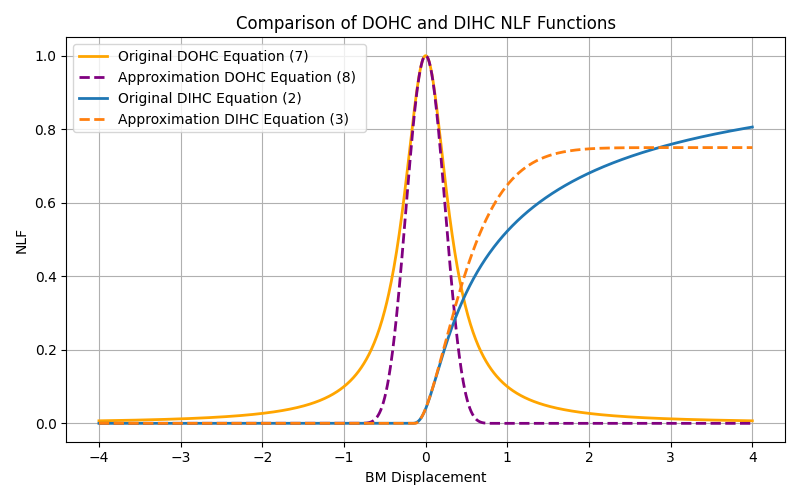}}
    \caption{The DIHC and DOHC NLF function approximation results}
    \label{fig:OHCIHCNLF}
\end{figure}
\begin{figure}[H]
    \centering
    \centerline{\includegraphics[height=4.6cm, keepaspectratio]{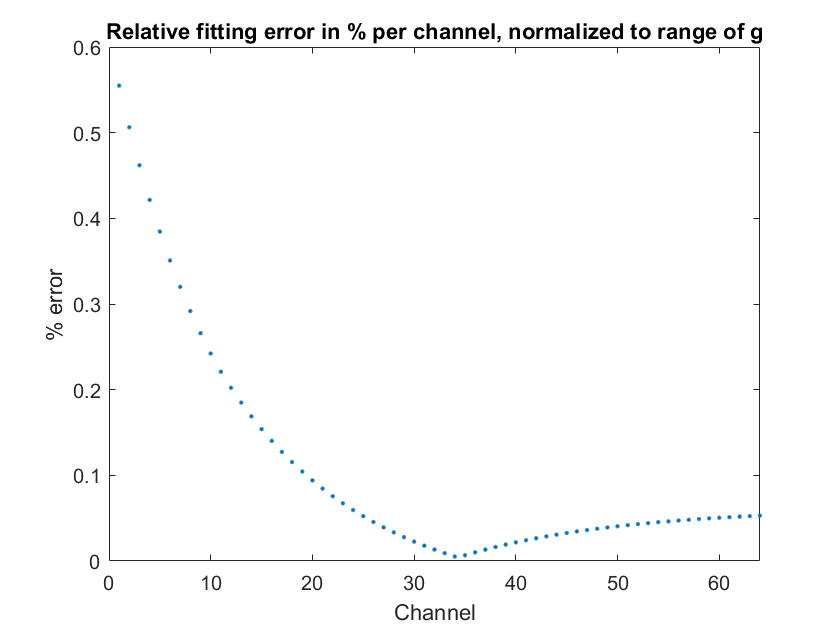}}
    \caption{Maximum fitting error between equation \ref{eq:g} and equation \ref{eq:gUpdate}}
    \label{fig:gApproximationError}
\end{figure}
The second division operation is in the DOHC: 
\begin{equation}
\text{sqr} = \left( \text{scale} \cdot v_{\text{OHC}} + \text{offset} \right)^2
\label{eq:sqr}
\end{equation}

\begin{equation}
\text{NLF} = \frac{1}{1 + \text{sqr}}
\label{eq:nlf}
\end{equation}
where the NLF models the DOHC compression function of the BM velocity, $v_\text{ohc}$, with a scale of 0.1 and an offset of 0.04. 
In our implementation, the DOHC NLF function is modeled by: 
\begin{equation}
\text{NLF} = \left[ \max\left(0,\ 1 - \frac{\text{sqr}}{8} \right) \right]^8
\label{eq:nlf_clip}
\end{equation}
The comparisons of the two NLFs approximations are shown in Fig. \ref{fig:OHCIHCNLF}. Both DIHC and DOHC NLF approximations are closely matched to the original equations when the BM displacement is small. As the BM displacement increases, for example, the DIHC approximation saturates more rapidly, which could limit the ability to represent amplitude at high levels.

The third division operation, as part of the $H$ of the BM in Fig. \ref{fig:SimplCARFAC} left bottom, is used to control the DC gain for the BM in equation \ref{eq:g}:
\begin{equation}
    g=\frac{1 - 2a_0 r + r^2}{1 - (2a_0 - h c_0) r + r^2}
    \label{eq:g}
\end{equation}
where $a_0$, $c_0$, and $h$ are the resonator coefficients and $r$ is the pole/zero radius in the $z$ plane:

\begin{equation}
    r = r_1 + d_\text{rz} \cdot \text{$u$}
    \label{eq:r}
\end{equation}
where $r_1$ is the pole/zero radius at maximum damping (widest filter), $d_\text{zr}$ is the range of variation controlling the radius increase as damping decreases, and $u$ is the undamping factor in [0,1] controlling damping reduction via DOHC feedback. The DC gain $g$ is monotonic and bounded within [0,1], allowing approximation by a second-degree polynomial in $u$:
\begin{equation}
    g = A u^2 + B u + C
    \label{eq:gUpdate}
\end{equation}
\begin{figure}
    \centering
    \centerline{\includegraphics[width=0.95\linewidth]{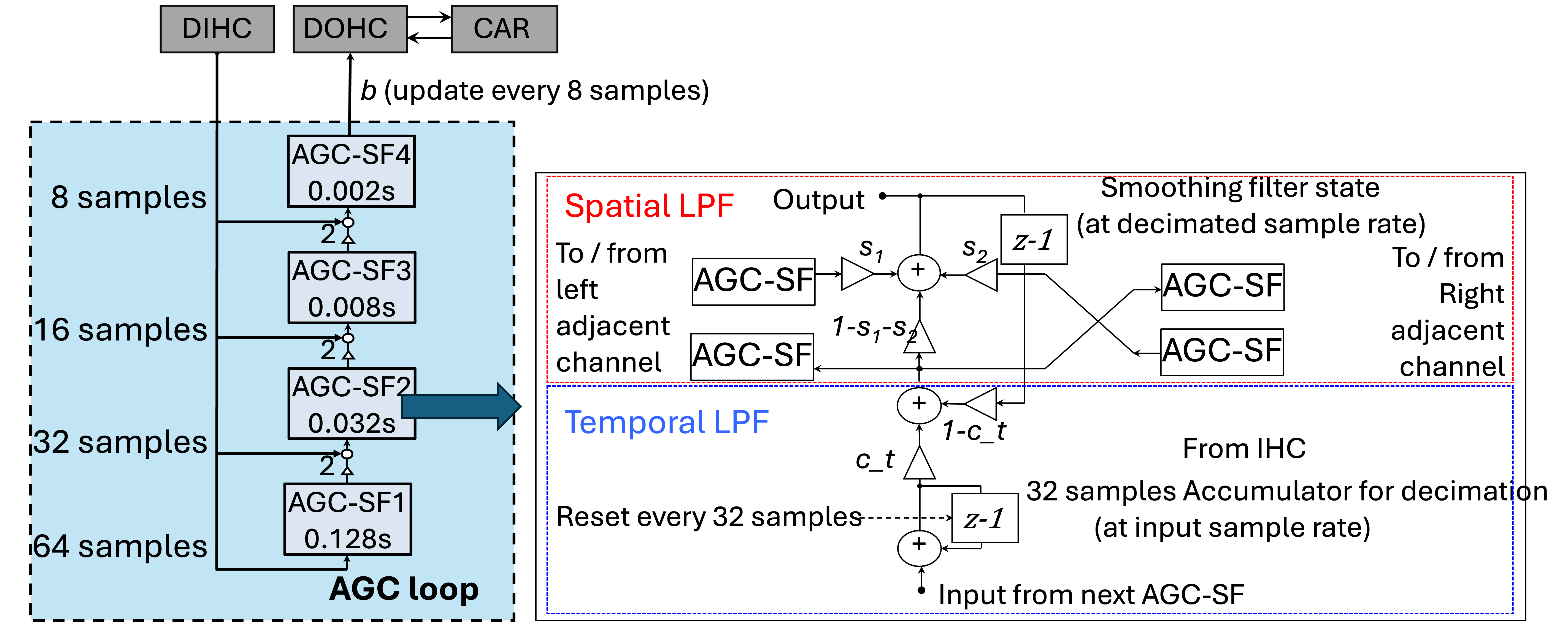}}
    \caption{Structure of the AGC loop. Four stages of the temporal smoothing filters (SF) . Each stage consists of a temporal LPF with a defined time constant (0.002, 0.008, 0.032, and 0.128 s) and a three-tap spatial smoothing filter. The internal structure of an AGC-SF, the input of the AGC-SF comes from the lower filter stage with the smaller time constant as well as the accumulation of the DIHC. The output goes to the next stage of the temporal filter. The spatial smoothing filter is a three-tap smoothing filter coupled with lateral channels. $s_1$, $s_2$, and $1-s_1-s_2$ are the spatial filter coefficients. $c_t$ is the temporal LPF coefficient calculated from the time constant.}
    \label{fig:AGC}
\end{figure}
where $A$, $B$ and $C$ are parameters that are pre-calculated lookup tables indexed by $u$ to approximate the division function of $g$. 
This approximation achieves a maximum error of 0.6\% when fitted to equation \ref{eq:g}, as shown in Fig. \ref{fig:gApproximationError}. 

The division replacements allow efficient real-time computation without computing the full rational expressions in hardware, particularly for the approximation of $g$, in which the numerator and the denominator vary over a wide range. It reduced the number of clock cycles required to 1, while only using up to two DSP slices. This approximation was developed under the guidance of Lyon for the purpose of hardware optimization and was also adopted in CARFAC v2 in \cite{Lyon2024TheJAX}.
\subsection{Time-multiplexing and Pipelining }
The cascade structure of the CARFAC model makes it well-suited for hardware implementation using time-multiplexing. In \cite{ Xu2018AModel}, a single H (CAR), DIHC, DOHC, and AGC loop was scaled through time-multiplexing to implement a multi-channel CARFAC, with BlockRAMs storing each channel’s parameters to enable circuit reuse. Building on this, we further optimized the time-multiplexing by using a single AGC temporal-spatial filter to construct the multi-time-scale AGC-loop, as shown in Fig. \ref{fig:AGC}. 

\begin{figure}
    \centering
    \centerline{\includegraphics[height=3.5cm, keepaspectratio]{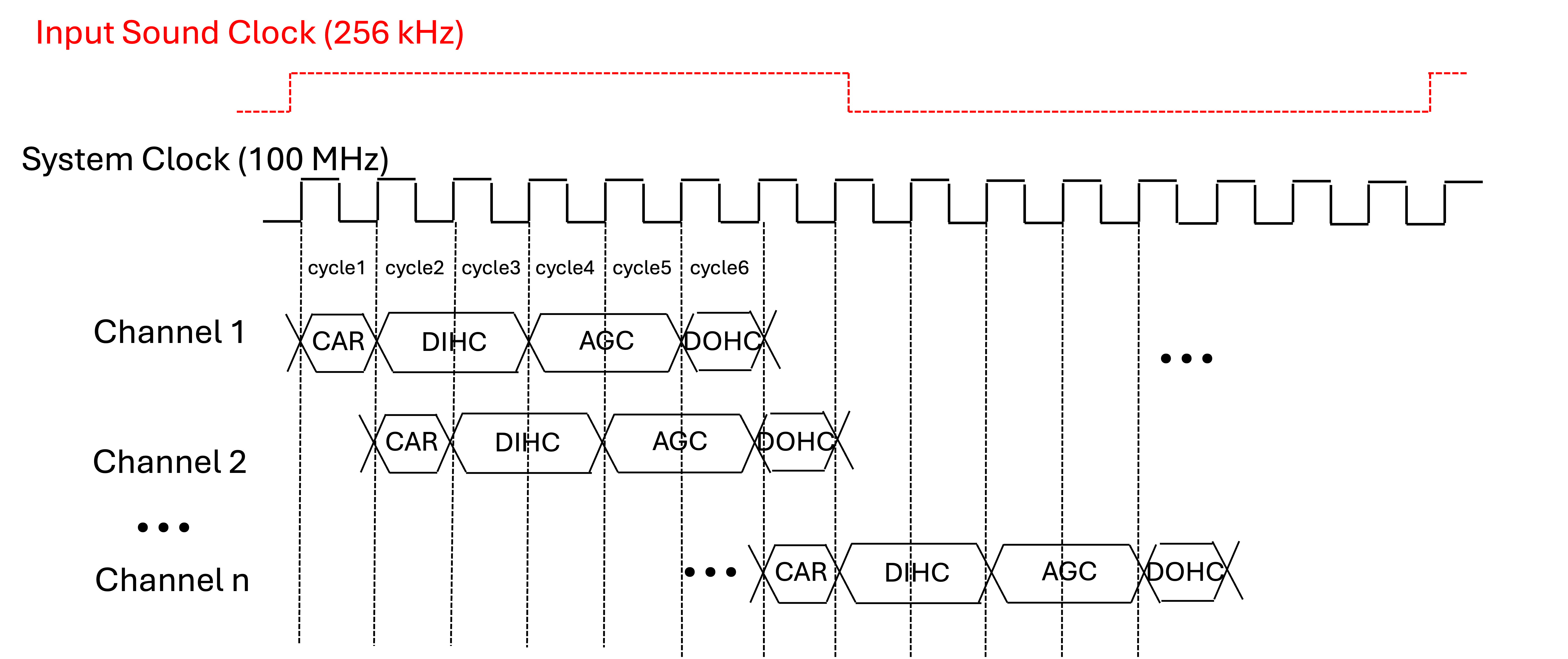}}
    \caption{The module-level time-multiplexing and pipeline diagram of CARFAC. The system clock operates at 100 MHz.}
    \label{fig:pipeline}
\end{figure}

The fast-acting compression of CARFAC is realized by updating $r$ through the DOHC. It includes instantaneous velocity feedback and multi-time-scale AGC\_loop feedback (via DIHC), as shown in Fig. \ref{fig:SimplCARFAC}. Due to the computational complexity of this loop, it is not feasible to complete this calculation within a single clock cycle. We implement this as a six-cycle pipeline, using clock-rate pipelining to distribute the operation across multiple system clock cycles. 
Delay balancing is applied to ensure that signals from different pipelines arrive simultaneously. In total, there are approximately 100 pipeline stages distributed across the CAR, DIHC, AGC, and DOHC modules. As the pipeline is too deep for low-level illustration, a module-level abstraction is shown in Fig. \ref{fig:pipeline}. 

In this work, time-multiplexing and pipeline paralleling are combined to minimize resource utilization while maintaining high throughput. 
\subsection{Quantization}
Floating-point operations are costly in hardware, so we converted the model to fixed-point. Signal resolution is limited by the 24-bit input, and most multiplications use 18-bit operands to fit DSP slice constraints. Additions preserve precision with an extra bit, and multiplications by two are replaced with bit shifts. The system was converted section-by-section, with numerical accuracy carefully monitored throughout. 
\section{Result}
The system was tested on an AMD Kria KV260 SoM, processing hydrophone (OceanSonics) data at a 256 kHz sampling frequency. 
The interface for feeding data from the external hydrophone is built on the embedded processor in Rust. 
Data transfer between the processor and accelerator is handled via a Direct Memory Access (DMA)-enabled AXI4-Stream bus. The accelerator is fully AXI4-Stream compatible and executes only when valid data is available, allowing the system to be used in real time with minimal latency. By using Rust, we achieve low resource usage while benefiting from the language’s modern design and memory safety guarantees. 
\begin{table}[H]
    \caption{Hardware usage of the complete design after synthesis and implementation in Vivado}
    \centering
    \includegraphics[height=3.5cm, keepaspectratio]{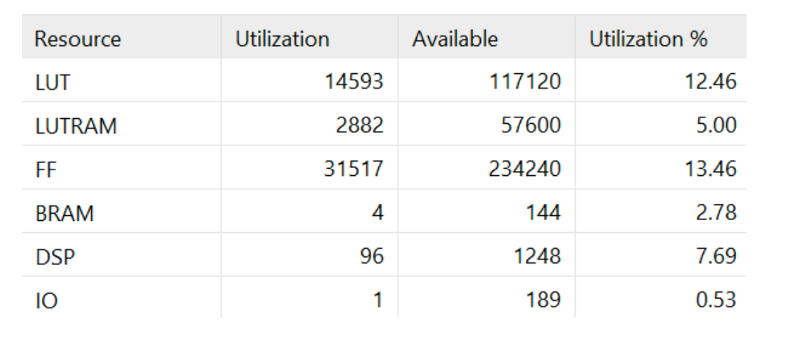}
    \label{tab:hardware-usage}
\end{table}
The accelerator is designed using MathWorks HDL Coder. This tool allows the graphical construction of a high-level representation, abstracting some of the building blocks typically required in hardware description languages. This approach not only reduces development time but also improves the reconfigurability and ease of validation. 
The system is synthesized using Vivado, at a 100 MHz system clock,  resulting in resource usage as shown in Table \ref{tab:hardware-usage}. 
The design only uses a maximum of 13.5\% of any of the resources,  which allows up to seven parallel CARFAC instances without further optimization.
We measured the power consumption of the Kria KV260 SoM during the execution of the system over a 15-minute window for the configuration of a 64-channel CARFAC, resulting in an average power draw of 3.11 W of the whole board. Meanwhile, the average load on the four ARM cores was measured to be below 7\%. 

\section{Acknowledgment}
The authors gratefully acknowledge the contributions of Richard Lyon for modeling the CARFAC in hardware and for his assistance in editing the manuscript.
\section{Conclusion and Discussion}
This paper presents a hardware-accelerated implementation of the CARFAC cochlea model on an embedded system. We improved the previous time-multiplexed and pipelined architecture and replaced costly division operations. The design provides an efficient platform for bio-inspired underwater acoustic sensing. CARFAC preserves fine temporal and phase details, which has the potential to integrate with beamforming techniques for robust underwater sound source localization. The system can be further extended to neuromorphic acoustic applications, as demonstrated in \cite{10558113, Xu2023}.

\bibliographystyle{ieeetr}       
\bibliography{references}
\end{document}